\documentclass[journal]{IEEEtran}
\usepackage{amssymb,amsmath,subfigure}
\usepackage{algorithm}
\usepackage{amssymb}
\usepackage{amsthm}
\usepackage{algpseudocode}
\usepackage{epsfig,amssymb,amsbsy,verbatim,array,caption,subfigure}
\newtheorem{definition}{Definition}
\newtheorem{note}{Note}

\begin{document}
\title{Self-Triggered Control in Artificial Pancreas}
\author{Debayani Ghosh,~\IEEEmembership{Member,~IEEE,}
        Sahaj Saxena,~\IEEEmembership{Member,~IEEE}
       and~Navin Kumar
\thanks{Debayani Ghosh is with the Department
of Electronics and Communication Engineering, Thapar Institute of Engineering and Technology, Patiala-147004, Punjab, India, e-mail: debayani.ghosh@thapar.edu.}
\thanks{Sahaj Saxena is with the Department
of Electrical and Instrumentation Engineering, Thapar Institute of Engineering and Technology, Patiala-147004, Punjab, India, e-mail: (sahaj.saxena@thapar.edu).}
\thanks{Navin Kumar is with the Department
of Mechanical Engineering, Indian Institute of Technology Ropar, Rupnagar-140001, Punjab, India, e-mail: nkumar@iitrpr.ac.in}
\thanks{Manuscript received August 15 202}}

\markboth{Preprint}%
{Shell \MakeLowercase{\textit{et al.}}: Bare Demo of IEEEtran.cls for IEEE Journals}
\maketitle
\begin{abstract}
The management of type 1 diabetes has been revolutionized by the artificial pancreas system (APS), which automates insulin delivery based on continuous glucose monitor (CGM). While conventional closed-loop systems rely on CGM data, which leads to higher energy consumption at the sensors and increased data redundancy in the underlying communication network. In contrast, this paper proposes a self-triggered control mechanism that can potentially achieve lower latency and energy efficiency. The model for the APS consists of a state and input-constrained dynamical system affected by exogenous meal disturbances. Our self-triggered mechanism relies on restricting the state evolution within the robust control invariant of such a system at all times. To that end, using tools from reachability, we associate a safe time interval with such invariant sets, which denotes the maximum time for which the invariant set remains invariant, even without transmission of CGM data at all times.
\end{abstract}

\IEEEpeerreviewmaketitle
\section{Introduction}
\IEEEPARstart{T}{ype} 1 diabetes is one of the concerning metabolic disorders where the $\beta$\textemdash cells of the pancreas in the patient are unable to produce insulin, thereby leading to an increase in blood glucose (glycemic) level. A closed-loop engineering device called an artificial pancreas offers its treatment by supplying insulin (through an insulin infusion pump) subcutaneously based on the glycemic measurement (through a continuous glucose monitoring system). Therefore, the delivery of insulin effectively and efficiently can be considered as one of the exciting problems for control engineers; the comprehensive details in APS engineering can be found in  \cite{kovatchev2019century,boughton2019advances} and the various control strategies in \cite{chee2007closed,mughal2024comprehensive,quiroz2019evolution,bhonsle2020review}.

In traditional control systems, sensor data is transmitted to the controller at regular time intervals. This can increase energy consumption and traffic congestion (data redundancy) in the context of the networked control system. Recently, two control paradigms have received widespread attention in the literature\textemdash event-and self-triggerred control \cite{heemels2012introduction,mazo2008event}. In these the exact sensor measurements are transmitted to the controller only when it is necessary. In event-triggered control, the sensor needs to monitor the state evolution continuously. However, in the self-trigger mechanism, the next transmission instant (scheduling) is specified at the current transmission instant. Hence, during this interval, the sensor can operate in sleep mode and considerably reduce energy consumption on the device's side. 

Motivated by this, we introduce, for the first time, the concept of a self-triggered scheduler in the APS. In APS, CGM usually transmits the glucose measurement periodically.  In contrast, we, by means of our proposed self-triggered mechanism, can make the control mechanism aperiodic, thereby improving the latency and reducing the transmissions, and achieving an energy-efficient system. 

Our model consists of a constrained dynamical system acted upon by an exogenous disturbance. Regarding APS, this translates to a model of the insulin-glucose relationship with unannounced meals as disturbances.
Specifically, the contributions of the work are as follows:
\begin{enumerate}
\item We identify the safe operating regime for APS in terms of a robust control invariant set of the insulin-glucose model. 
    \item We propose a novel self-triggered mechanism for APS, which yields a set of feasible scheduling sequences which specify the exact time instants at which the CGM can send measurements to the controller for computation of control actions (insulin rate). Our self-triggered mechanism also guarantees the safe operation of APS at all times thereby enhancing its reliability. 
    \item Our proposed scheduler can achieve less energy consumption at the device side considerably.
   \end{enumerate}
The structure of this brief is as follows. We describe the preliminaries and the system model of APS  in Section II. In Section III, the problem is formulated.  The effectiveness of the proposed strategies is illustrated by simulation results in Section IV. This brief is concluded in Section V.

\section{System Model and Preliminaries}
In this work,  we consider the patient model for APS as the dynamics of insulin-glucose relation, as deviation from an input of basal rate and output of 110 mg/dL \cite{van2011control}.  The model is described by a discrete linear time-invariant system affected by exogenous disturbances (unannounced meal):
\begin{equation}
\begin{gathered}
x_{t+1}=Ax_t+Bu_t+Ew_t,\\ 
y_t=Cx_t,   
\label{eq:discrete_sys}
\end{gathered}
\end{equation}
where $x_t$ is the state and is subject to a  polytopic constraint of the form: $x_t\in\mathcal{X}\subseteq \mathbb{R}^{n},$ $u_t$ is the control input subject to the constraint $u_t\in\mathcal{U}\subseteq \mathbb{R}$, and $w_t$ is the external disturbance and subject to the following polytopic constraint $w_t\in\mathcal{W}\subseteq \mathbb{R}.$ Further, $A\in\mathbb{R}^{3\times 3}, B\in\mathbb{R}^{3\times 1}, E\in\mathbb{R}^{3\times 1}\,\,\text{and}\,\,C\in\mathbb{R}^{1\times 3}.$
\begin{definition}
A control input $u_t$ for the system (\ref{eq:discrete_sys}) is said to be admissible if $u_t\in\mathcal{U}$.  
\end{definition}
\begin{definition}
A set $\mathcal{C}$ is said to be a robust control invariant set for system (\ref{eq:discrete_sys}) if for all  $x_t\in\mathcal{C}$, there exists an admissible control input $u(t)\in\mathcal{U}$ such that $x_{t+1}\in\mathcal{C}$, $\forall w_t\in \mathcal{W}$. That is, $\mathcal{C}$ is a control input invariant if we can find an admissible control input for all $x_t\in\mathcal{C}$ such that it keeps the next step state $x_{t+1}\in\mathcal{C}$, irrespective of the realized exogenous disturbance.
\end{definition}

\begin{note}
Such an invariant set can be computed using tools from MATLAB's multi-parametric toolbox (MPT) \cite{MPT3}.
\end{note}
\begin{definition}
 A set $\mathcal{C}_{\infty}\subseteq\mathcal{X}$ is said to be a maximal robust control invariant set for the system (\ref{eq:discrete_sys}) if it contains all control invariant sets contained in $\mathcal{X}$. 
\end{definition}
\begin{note}
For a polytopic state space $\mathcal{X}$, the maximal control invariant set will also be a polytope that admits a $H-$representation \cite{borrelli2017predictive}.      
\end{note}

\begin{note}
The external disturbance $w_t$ in the APS model \eqref{eq:discrete_sys} can also be thought of as a fault injection or a cyber-attack.
\end{note}

\section{Problem Formulation}
Given a maximal invariant set $\mathcal{C}$ of the system (\ref{eq:discrete_sys}), we aim to devise a self-triggered scheme for the system (\ref{eq:discrete_sys}). To achieve this, we first associate a safe time interval with the invariant set, using reachability tools outlined in \cite{borrelli2017predictive}. We then identify that the invariant set would remain invariant for more than one step, and is governed by the safe time interval. We now outline our proposed methodology for devising the same.

Suppose $x_0\in\mathcal{C}_{\infty}$ and CGM measurements are transmitted at $t=0$. For $t>0$, we assume that there is no further communication from the sensor to the controller. This implies that the controller does not have access to the actual state measurements for $t>0.$ For any time-step $j$, we first define the feasible sets $X_j$ as 
 \begin{align}
     X_j=&\lbrace x_0\in\mathcal{C}_{\infty},~\exists~~ \lbrace u_0,u_1,\hdots u_{t-1}\rbrace\in\underbrace{\mathcal{U}\times\mathcal{U}\times\dots\mathcal{U}}_{t \,\,\text{times}},\notag\\ &\textrm{s.t.}\,\,x_t\in\mathcal{C}_{\infty},\,\forall w_i\in \mathcal{W}\,\,\forall\,\,t\in\lbrace 1,2,\hdots,j\rbrace\}.
     \end{align}
Then, we mathematically define the safe time interval associated with the maximal invariant set as the following:
 \begin{align}
     \alpha=&\max\{j~|~X_j=\mathcal{C}_{\infty}\}.
     \end{align}
We call $\alpha$ as the safe time interval associated with the maximal interval set $\mathcal{C}_{\infty}$. In other words, $\alpha$ is that time interval for which the maximal invariant set remains invariant without further communication after $t=0$ under some admissible inputs. To calculate $\alpha$, we adopt the following approach:
\noindent Suppose the $H-$representation of the maximal invariant set is 
\begin{equation}
    Hx\le h, 
\end{equation}
and the $H-$representation of the input constraints is 
\begin{equation}
    H_u u\le h_u.
\end{equation}
We now observe that
\begin{align}
x_1&=Ax_0+Bu_0+Ew_0,\notag\\
x_2&=Ax_1+Bu_1+Ew_1\notag\\
  &=A[Ax_0+Bu_0+Ew_0]+Bu_1+Ew_1\notag\\
  &=A^2x_0+ABu_0+Bu_1+AEw_0+Ew_1,\notag\\
&\vdots\notag\\
x_j&=A^jx_0+A^{j-1}Bu_0+A^{j-2}Bu_1+\ldots+Bu_{j-1}\notag\\&+A^{j-1}Ew_0+A^{j-2}Ew_1+\ldots+Ew_{j-1}. 
\end{align}
\begin{note}
From (5),  we note that the state at $j$-th instant can be denoted only in terms of $x_0$ and the sequence of control inputs $u_0, u_1, \ldots,u_{j-1}$.    
\end{note}

To calculate the safe time interval $\alpha$, we see that the following constraints should hold for any time step $j$: 
\subsection*{C1: State Constraints}
$x_0\in\mathcal{C}_{\infty}$, $x_t\in\mathcal{C}_{\infty}$  for all $t\in{1,2,\ldots,j}$, 
\begin{equation}
\begin{gathered}
    Hx_0 \le h\\
    Hx_1 \le h\\
    \vdots\\
    Hx_j \le h.
\end{gathered}
\end{equation}

\subsection*{C2: Input Constraints}
The sequence of control inputs $u_0,u_1,\ldots,u_{j-1}$ should be admissible, i.e., 
$u_t\in\mathcal{U}_{\infty}$  for all $t\in\left\{0,1,\ldots,j-1\right\}$, 
\begin{equation}
\begin{gathered}
    H_uu_0\le h_u\\
    \vdots\\
    H_uu_{j-1}\le h_u
\end{gathered}
\end{equation}

\noindent Stacking constraints (6) and (7) together and using (5) we have the following system of matrix inequalities:
\begin{equation}
 M^{j} \begin{pmatrix}
    x_0 \\
    \hat{u}
  \end{pmatrix}\leq P^{j}-G^{j}\hat{w}
  \label{eq:myeqn}
\end{equation}
where $\hat{u}=[u_0, u_1,\ldots,u_{j-1}]^T$. $M^j$ and $p^j$ can be found by stacking constraints (6) and (7). To incorporate the effect of the worst-case disturbance, we tighten the constraints in $(8)$ by solving a maximization problem for each row of $G^j.$ This reduces to a simple linear problem, since $\mathcal{W}$ is a polytope, and can be easily solved using MPT. The resulting polyhedron contains all possible solutions for $x_0$ and its corresponding input sequence $\hat{u}$ irrespective of the disturbance. We project (8) onto its first $3$ coordinates (dimension of the system), giving the set of initial states for which the future state evolution for $j$ steps lie in $\mathcal{C}_{\infty}$. Then, the procedure for finding $\alpha$ can be outlined algorithmically below:
    \begin{algorithm}
    \caption{Algorithm for computation of safe time}\label{euclid}
    \hspace*{\algorithmicindent} \textbf{Input:} Maximal Invariant Set $\mathcal{C}_{\infty}$\\
    \hspace*{\algorithmicindent} \textbf{Output:} Safe time interval $\alpha$ 
    \begin{algorithmic}[1]
    \For {$n=1:j$}
    \State Form the system of matrix inequalities (8)
    \State Project (8) onto first $3$ coordinates
    \State $C^j$ = projection$\left\{y^3~|~ M^j\begin{bmatrix}
        x_0\\\hat{u}
    \end{bmatrix}\le P^{j}-G^{j}\hat{w}\right\}$
    \If {$\mathcal{C}_j=\mathcal{C}_{\infty}$}
   \State Continue
    \Else
    \State Break
    \EndIf
    \State $\alpha=j$
    \EndFor
    \end{algorithmic}
    \end{algorithm}

\subsection{Formation of Scheduling}
Given the safe interval $\alpha$, we can identify that the CGM measurement should be transmitted at least once in the duration of $\alpha$ time slots. Hence, any such transmission will give rise to a feasible scheduling sequence. For example, if $\alpha$ is identified to be 3, then some examples of valid scheduling sequences can be 
\begin{align*}
 \Lambda_1&=\lbrace 0,2,4,7,10,\cdots\rbrace,\\
 \Lambda_2&=\lbrace 0,3,6,9,12,\cdots\rbrace.  
\end{align*}

\begin{figure}
    \centering
    \includegraphics[width=\linewidth]{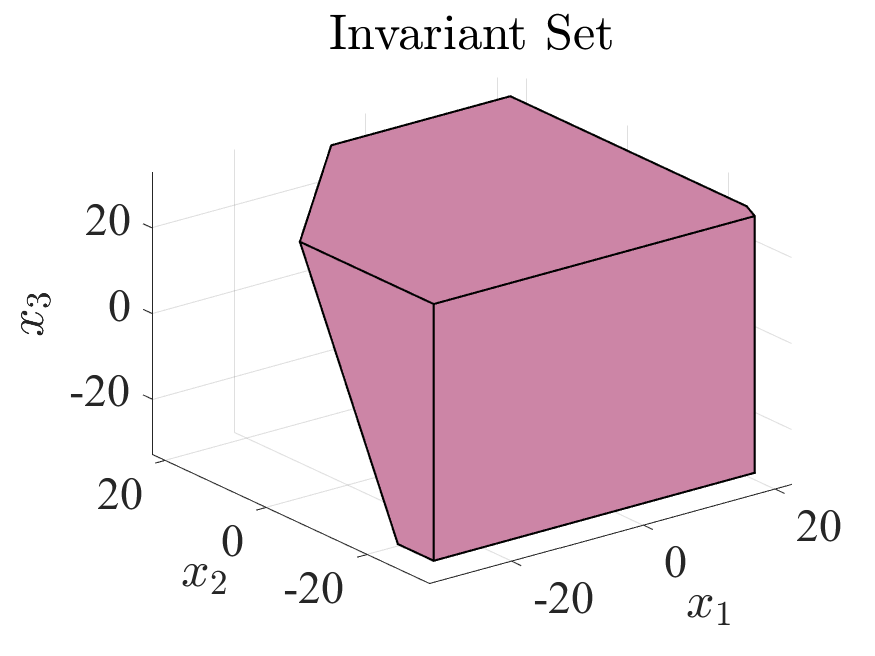}
    \caption{Invariant set for insulin-glucose model}
    \label{fig:invariant set}
\end{figure}
\begin{figure}
 \begin{center}
  \subfigure[]{
  \label{fig:buffer_kappa}
  \includegraphics[width=\linewidth]{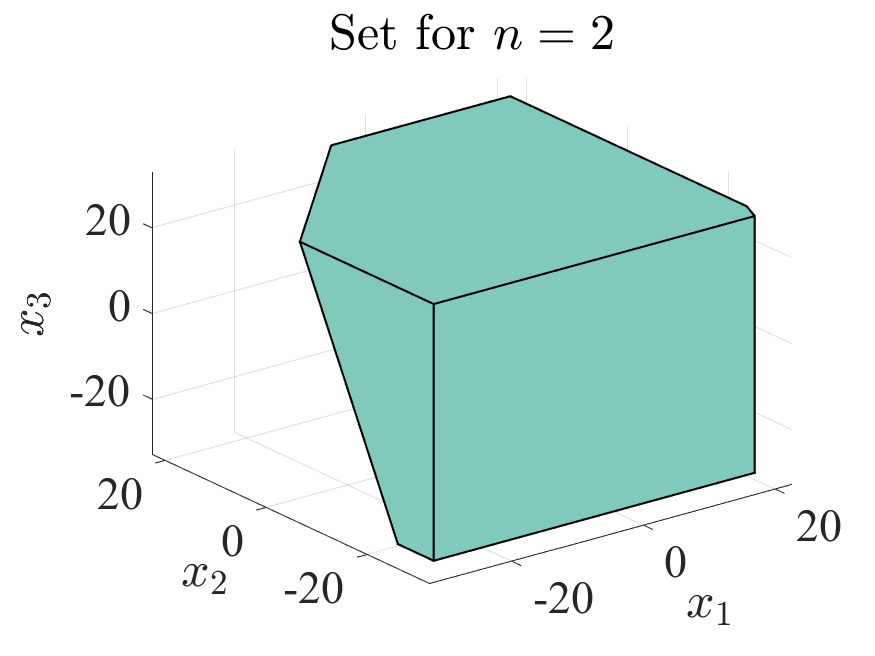}
  }
    \subfigure[]{
  \label{fig:buffer_kappa}
  \includegraphics[width=\linewidth]{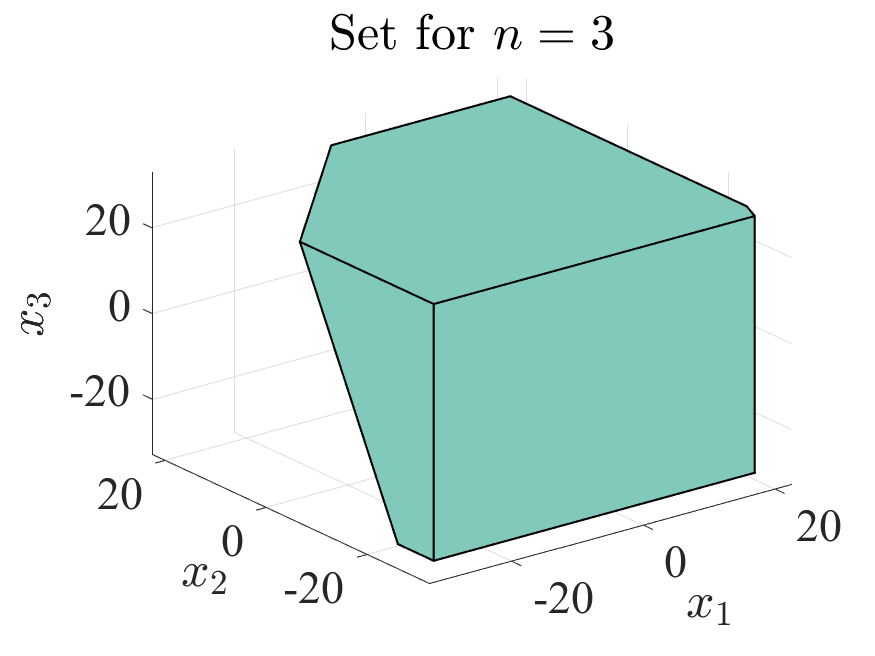}
  }
    \subfigure[]{
  \label{fig:buffer_kappa}
  \includegraphics[width=\linewidth]{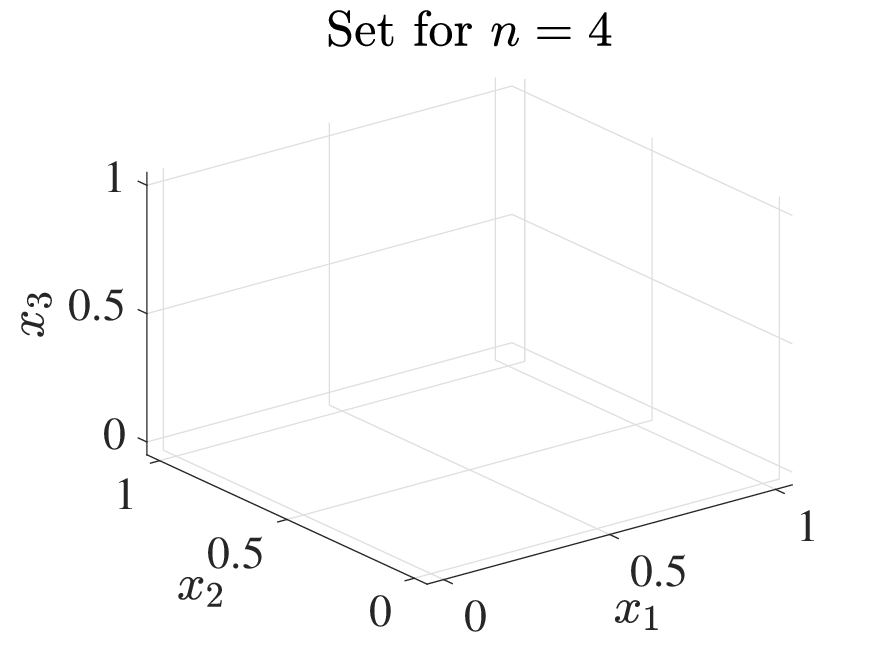}
  }
\caption{Solution of the set of initial states $x_0$ obtained from Algorithm 1 for (a)  $n=2$ (b) $n=3$ and (c) $n=4$.}
  \label{fig:charts}
 \end{center}
\end{figure}
\section{Results and Discussions}
To establish the correctness of our proposed algorithm, we conduct extensive simulations in the MPT in MATLAB. We consider the following APS model as outlined in Section II with the following matrices:
\begin{align*}
   A = \begin{bmatrix}
-a_1 & -a_2 & -a_3\\
1 & 0 & 0\\
0 & 1 & 1
\end{bmatrix}, 
B = \begin{bmatrix}
K \\ 0 \\ 0 
\end{bmatrix},
E = &\begin{bmatrix}
0 \\ 0 \\ 1 
\end{bmatrix},\\
 C = \begin{bmatrix}
0 & 0 & 1 
\end{bmatrix}.
\end{align*}
Further, we consider the following values: 
\begin{align*}
 K &= -2,\\
 a_1 &= -0.965\times2-0.98,\\
 a_2 &= 2\times0.98\times0.965+0.965^2\\
 a_3 &= -0.98\times0.965^2
\end{align*}
We also consider the following constraints on the input insulin rate as $$-10\leq u_t\leq 100,$$ and constraints on the state as $$-30\leq x_t\leq 30.$$ Further, we consider the constraints on the exogenous disturbance as $$0\leq w_t\leq 10.$$

We first compute the maximal invariant set of the system using the algorithm outlined in \cite{borrelli2017predictive}. The invariant set is shown in Fig. \ref{fig:invariant set}. Then, using our proposed Algorithm $1,$ we compute the safe time interval $\alpha$ associated with the maximal invariant set. Our simulation results in Fig. 2 show that the invariant set $\mathcal{C}_{\infty}$ remains invariant for 3-time steps, given the CGM transmits exact sensor measurements at $t=0.$ This implies that given a sensor data transmission at $t=0,$ we have a 3-step control input sequence $\lbrace u_0,u_1,u_2\rbrace$ which restrict the state evolution within the safe set, \emph{i.e.}, the invariant set for the next 3 time steps. Within these 3-time steps, the CGM data can choose not to send the exact measurements to the controller and can essentially go to sleep. Hence, according to our algorithm, the safe time interval associated with the APS system is $\alpha=3.$

Note that to minimize energy consumption at the CGM, our proposed self-triggered mechanism may choose to transmit the data to the APS at intervals of exactly $3$ time-steps. This would lead to a $67.67\%$ reduction in energy consumption as compared to a periodic APS, which would typically monitor and transmit the sensor data at all time steps. 

\section{Conclusions}
This work proposes a  self-triggered control mechanism for the APS in the management of type 1 diabetes. By minimizing the reliance on continuous glucose monitoring data, this approach offers the potential for reduced energy consumption and lower latency in insulin delivery. The self-triggered mechanism relies on restricting the system states within an invariant set, thus ensures that glucose levels remain within safe boundaries, even in the presence of meal disturbances, without the need for constant CGM data transmission. Our work not only addresses the limitations of conventional closed-loop systems but also paves the way for more efficient and reliable diabetes management solutions. 
\section*{Acknowledgment}
This work is funded by the Science and Engineering Research Board, Government of India, under the Tare Scheme (TAR/2021/000297).
\bibliographystyle{IEEEtran}
\bibliography{Ref}

\begin{thebibliography}{10}
\providecommand{\url}[1]{#1}
\csname url@samestyle\endcsname
\providecommand{\newblock}{\relax}
\providecommand{\bibinfo}[2]{#2}
\providecommand{\BIBentrySTDinterwordspacing}{\spaceskip=0pt\relax}
\providecommand{\BIBentryALTinterwordstretchfactor}{4}
\providecommand{\BIBentryALTinterwordspacing}{\spaceskip=\fontdimen2\font plus
\BIBentryALTinterwordstretchfactor\fontdimen3\font minus
  \fontdimen4\font\relax}
\providecommand{\BIBforeignlanguage}[2]{{%
\expandafter\ifx\csname l@#1\endcsname\relax
\typeout{** WARNING: IEEEtran.bst: No hyphenation pattern has been}%
\typeout{** loaded for the language `#1'. Using the pattern for}%
\typeout{** the default language instead.}%
\else
\language=\csname l@#1\endcsname
\fi
#2}}
\providecommand{\BIBdecl}{\relax}
\BIBdecl

\bibitem{kovatchev2019century}
B.~Kovatchev, ``A century of diabetes technology: signals, models, and
  artificial pancreas control,'' \emph{Trends in Endocrinology \& Metabolism},
  vol.~30, no.~7, pp. 432--444, 2019.

\bibitem{boughton2019advances}
C.~K. Boughton and R.~Hovorka, ``Advances in artificial pancreas systems,''
  \emph{Science Translational Medicine}, vol.~11, no. 484, p. eaaw4949, 2019.

\bibitem{chee2007closed}
F.~Chee and T.~Fernando, \emph{Closed-loop control of blood glucose}.\hskip 1em
  plus 0.5em minus 0.4em\relax Springer Science \& Business Media, 2007, vol.
  368.

\bibitem{mughal2024comprehensive}
I.~S. Mughal, L.~Patan{\`e}, and R.~Caponetto, ``A comprehensive review of
  models and nonlinear control strategies for blood glucose regulation in
  artificial pancreas,'' \emph{Annual Reviews in Control}, vol.~57, p. 100937,
  2024.

\bibitem{quiroz2019evolution}
G.~Quiroz, ``The evolution of control algorithms in artificial pancreas: A
  historical perspective,'' \emph{Annual Reviews in Control}, vol.~48, pp.
  222--232, 2019.

\bibitem{bhonsle2020review}
S.~Bhonsle and S.~Saxena, ``A review on control-relevant glucose--insulin
  dynamics models and regulation strategies,'' \emph{Proceedings of the
  Institution of Mechanical Engineers, Part I: Journal of Systems and Control
  Engineering}, vol. 234, no.~5, pp. 596--608, 2020.

\bibitem{heemels2012introduction}
W.~P. Heemels, K.~H. Johansson, and P.~Tabuada, ``An introduction to
  event-triggered and self-triggered control,'' in \emph{2012 51st IEEE
  Conference on Decision and Control}.\hskip 1em plus 0.5em minus 0.4em\relax
  IEEE, 2012, pp. 3270--3285.

\bibitem{mazo2008event}
M.~Mazo and P.~Tabuada, ``On event-triggered and self-triggered control over
  sensor/actuator networks,'' in \emph{2008 47th IEEE Conference on Decision
  and Control}.\hskip 1em plus 0.5em minus 0.4em\relax IEEE, 2008, pp.
  435--440.

\bibitem{van2011control}
K.~Van~Heusden, E.~Dassau, H.~C. Zisser, D.~E. Seborg, and F.~J. Doyle~III,
  ``Control-relevant models for glucose control using a priori patient
  characteristics,'' \emph{IEEE Transactions on Biomedical Engineering},
  vol.~59, no.~7, pp. 1839--1849, 2011.

\bibitem{MPT3}
M.~Herceg, M.~Kvasnica, C.~Jones, and M.~Morari, ``{Multi-Parametric Toolbox
  3.0},'' in \emph{Proc.~of the European Control Conference}, Z\"urich,
  Switzerland, July 17--19 2013, pp. 502--510,
  \url{http://control.ee.ethz.ch/~mpt}.

\bibitem{borrelli2017predictive}
F.~Borrelli, A.~Bemporad, and M.~Morari, \emph{Predictive control for linear
  and hybrid systems}.\hskip 1em plus 0.5em minus 0.4em\relax Cambridge
  University Press, 2017.

\end{thebibliography}

\end{document}